\documentclass[11pt,twoside]{article}

\usepackage{baaa-eng}
\usepackage{graphicx}
\usepackage{subfigure}
\usepackage{psfrag}

\usepackage{epsfig}

\usepackage[spanish]{babel}
\usepackage[latin1]{inputenc}

\usepackage[T1]{fontenc} 
\usepackage{ae,aecompl} 

\usepackage{latexsym}
\usepackage{verbatim}
\usepackage[colorlinks=true,dvips]{hyperref}

\def\etal{et\thinspace al.\thinspace}                    

\begin{document}
\vskip 1.0cm
\markboth{ R. Cid Fernandes }%
{Paleontology of Galaxies}

\pagestyle{myheadings}
\vspace*{0.5cm}
\parindent 0pt{INVITED PAPER            } 
\vskip 0.3cm

\title{ Paleontology of Galaxies: Recovering the Star Formation \&
Chemical Enrichment Histories from Galaxy Spectra}

\author{Roberto Cid Fernandes$^{1}$}

\affil{%
  (1) Depto. de F\'\i sica, Universidade Federal de Santa Catarina
  (Brasil)
}

\begin{abstract} 
Recent advances in stellar population modelling and avalanches of data
from mega-surveys have revived the interest in techniques to extract
information about galaxy evolution from integrated spectra. This
contribution provides an informal and (hopefully) pedagogical, but
inevitably biased and incomplete introduction to this field. Emphasis
is given to the several choices one has to make in the process of
modelling galaxy spectra.
\end{abstract}

\begin{resumen}
Los recientes avances en modelos de poblaciones estelares y la avalancha
de datos de mega-surveys han revivido el inter\'es en t\'ecnicas para
extraer informaci\'on acerca de la evoluci\'on de galaxias a partir de
espectros integrados. Esta contribuci\'on presenta una introducci\'on
informal y pedag\'ogica, pero inevitablemente sesgada e incompleta, a
este tema, con \'enfasis en las varias elecciones que uno tiene que
hacer en el proceso de modelar espectros de galaxias.
\end{resumen}

\section{Introduction}

The Wikipedia entry for ``paleontology'' states that it ``is the
study of the history and development of life on Earth, including that
of ancient plants and animals, based on the fossil record''. Much of
astronomical research shares this same general goal: To study how
things form and evolve, be they planets, stars, galaxies or the
universe as a whole.

A great deal of current work is focused on finding and characterizing
galaxies as far as possible.  Since far in space = far in time,
observations at different redshifts ($z$) sample different stages in
the history of galaxies. Studying how galaxy properties (or relations
between properties) change with $z$ gives us important clues on
evolutionary processes. The evolution of the mass-metallicity relation,
for instance, has now been extended to intermediate and high $z$'s
(e.g., Savaglio \etal 2005; Shapley \etal 2005), and comparison with
its local universe counterpart reveals systematic offsets which
ultimately reflect the history of metal enrichment along cosmic
time. Similarly, the (Tully-Fisher) relation between luminosity and
rotation speed has been mapped at different lookback times in search
of evolutionary effects (Arag\'on-Salamanca 2006). These are just two
examples of the thriving times of current observational cosmology.

Paleontologists would have a much easier life if they had a
``time-machine'' as nice as the universe! No more digging on dirty
swamps or hot deserts, no more trying to figure what a dinosaur
looked like from a couple of broken bones. Just find a few Earth like
places whose clocks mark different evolutionary times, and see how
things were and how evolution proceeded over the years.

Though promising, the high-$z$ work does not (yet) compete with local
universe studies in terms of data quality and specially
quantity. Limits on data quality, say low $S/N$ spectra, propagate to
limits in the amount of information that can be extracted about
galaxian histories. {\it Quantity} is equally important. Whereas just
a few years ago we would be happy with 2 or 3 digits samples,
mega-surveys like the 2dF and SDSS have taught us what a difference
$N$ makes!  Galaxy evolution is surely not a simple 1 or 2 parameter
problem. A helpful strategy to tackle complex multi-dimensional
problems is to break a sample in narrow ranges of physical or
observational properties (say, mass {\em and} environment {\em and}
color), which requires very large samples.  Kauffmann \etal (2003),
Brinchmann \etal (2004), Clemens \etal (2006), Mateus \etal (2006) and
Muriel's paper in this same volume are examples of studies which
benefited from $N \sim 10^{5\pm1}$ statistics.

But how can we dig the history of galaxies from cosmologically shallow
surveys, which barely reach a tenth of a galaxy's life?

\section{Fossil methods}

This is where the analogy with paleontology fits in. Faced with
fossils from different epochs, a paleontologist tries to figure, with
the help of ancillary data and/or theories, how things unfolded as
time passed. A galaxy spectrum can be seen as a {\em fossil record} of
its history. The old, low mass stars which shine today in elliptical
galaxies are just the long-lived relics of a hectic past, where stars
formed at high rates and massive stars dominated the scene. The gas
and just-born stars which make up today's HII regions in spirals
coexist with the low mass survivors of previous generations, whose
long-gone massive stars contributed to the metals now shining under
the ionizing photons of younger generations.

The age and metal content of young and old stars leave their
fingerprints as colors and absorption lines.  Hence, a galaxy's
spectrum contains information from {\it all} its stellar
generations. This is good news. The bad news is that unscrambling this
crazy mixture of photons coming from $\sim$ the same space but
reflecting different cosmic times is no easy task. The difficulties,
both mathematical and astrophysical, are best appreciated with some
basic formalism.

The most general strategy to tackle this puzzle is to consider a
galaxy as a sum of simple stellar populations (SSP), each with its age
($t_j$) and metallicity ($Z_j$), such that a galaxy spectrum can be
written as:

\begin{equation}
\label{eq:synt}
L_\lambda^{gal}(\vec{x}) = \sum_{j=1}^{N_\star} L^{SSP}_{\lambda,j} = 
L_{\lambda_0}^{gal} \sum_{j=1}^{N_\star} x_j l^{SSP}_{\lambda,j}
\end{equation}

\noindent where $l^{SSP}_{\lambda,j}(t_j,Z_j)$ is the spectrum of the
$j^{th}$ population normalized at a reference wavelength $\lambda_0$,
and $\vec{x}$ is the so called {\it population vector}, which says
what fraction of the galaxy light at $\lambda_0$ comes from the
$j^{th}$ population, i.e., from stars of age $t_j$ and metallicity
$Z_j$. Equivalently, one may express the problem in terms of {\it
mass} fractions, which relate to $\vec{x}$ via the $M/L_{\lambda_0}$
of each population. One then compares this model with an observed
spectrum $O^{gal}_\lambda$ (or indices derived from it) and seeks the
$\vec{x}$ which minimizes residuals:

\begin{equation}
\label{eq:chi2}
\chi^2 = 
\sum_\lambda [O^{gal}_\lambda - L^{gal}_\lambda(\vec{x})]^2 w_\lambda^2
\end{equation}

\noindent where $w_\lambda$ is the weight attributed to pixel
$\lambda$.  This equation is incomplete in at least two senses. First,
it lacks a term to account for the Doppler shifts caused by the way
stars move around in a galaxy, smoothing absorption lines on scales of
order 100 km$\,$s$^{-1}$. This effect is easily modeled convolving
equation \ref{eq:synt} with a kinematical filter, so let's pretend
this has been done. Secondly, extinction terms, which affect galaxy
spectra on a much larger $\lambda$-scale, are missing. Extinction can
and must be modelled, and the simplest thing to do is to model it as a
homogeneous foreground dust screen, multiplying \ref{eq:synt} by
$10^{-0.4 (A_\lambda - A_{\lambda_0})}$ and choosing some $A_\lambda /
A_{\lambda _0}$ reddening curve. Though this strategy gives useful
results, it entails a blatantly simplistic representation of real
galaxies, where different populations are affected differently by
dust. Accounting for this in a consistent and robust way is perhaps
the most serious challenge for current population synthesis
methods. In the absence of a solid technique to deal with this
difficulty, we are forced to adopt the traditional method of
forgetting the problem and adopting the screen model until further
notice.

Clearly, it is $\vec{x}$ we are after, as it encodes the distribution
of stellar ages and metallicities, and hence the whole star formation
history (SFH) of a galaxy.
To get there, we need to specify several things.

\subsection{The basis}

\label{sec:basis}

What should we use for $l^{SSP}_{\lambda,j}$? In other words,
which the {\it basis} in which we are going to decompose our model for
$L_\lambda^{gal}$?
The superscript ``SSP'' already tells what my favorite answer
is. Thinking of a galaxy spectrum in terms of a sum of SSPs is the
simplest reasonable thing one can think of. One then needs to decide
where will the basis spectra $l^{SSP}_{\lambda,j}$ ($j = 1 \ldots
N_\star$) be drawn from.

One approach, founded by Bica \& Alloin in the 80's, is to use {\it
observed star cluster spectra}. This strategy has proven quite
successful, and is still being actively pursued, as illustrated by the
work of Ahumada in this same volume. An advantage of using observed
clusters is that one mimics Mother Nature as best as possible, without
having to resort to assumptions about initial mass function, stellar
evolution and stellar spectra.  A disadvantage is that observed
clusters do not span the whole $t$-$Z$ space homogeneously (and curing
this deficiency is precisely one of the motivations of Ahumada's
work). Bica (1988) was the first to apply this technique to infer
galaxian SFHs.

Alternatively, one may use {\it models} for $l^{SSP}_{\lambda,j}$. The
modelling of SSP spectra has evolved so much that nowadays one can fit
observed cluster spectra close to perfection. Despite the larger load
on assumptions, theoretical $l^{SSP}_\lambda$'s can replace observed
ones with the advantage of a wider coverage of the $t$-$Z$ plane and
spectral range. Up to the publication of medium spectral resolution
($\lambda/\Delta\lambda \sim 2000$) models by Bruzual \& Charlot
(2003, BC03), soon followed by similar work by Le Borgne \etal (2004),
Gonzalez-Delgado \etal (2005) and others, this approach enjoyed
limited use. Since then, however, it has boomed. 

These are not the only alternatives. Instead of a summation of SSPs
one can model the SFH of galaxies in more continuous terms, adjusting
equation \ref{eq:synt} to an integral and so on.  An exponentially
declining star-formation rate function, for instance, is a quite
popular recipe. Instantaneous or finite duration bursts may be
superimposed to this continuous regime, and in fact any other {\it ad
hoc} or physically motivated parameterization may be implemented. As
long as one allows for many SSPs in \ref{eq:synt} (i.e., $N_\star \gg
1$), the discrete SSP-based approach encompasses more continuous
descriptions of galaxy histories. Given the many things which can spur
star formation throughout a galaxy's life, a {\it non-parametric}
approach sounds more advisable.

\subsection{Observables: Indices $\times$ Full Spectrum}

\label{sec:observables}

Having decided what spectral basis to use, you must specify your {\it
observables}, ie, what it is exactly that you will try to fit.  Eq.\
\ref{eq:synt} is written to model the pixel-by-pixel spectrum of a
galaxy, clearly the most complete approach possible. Historically,
however, astronomers have most often chosen to model not the full
$L^{gal}_\lambda$, but {\it spectral indices} such as absorption line
equivalent widths and/or colors. Indices can be seen as a compressed,
but highly informative, representation of the whole spectrum.

A particularly influential set of absorption line indices has been
defined by the Lick Observatory group on the basis of low resolution
spectra of a library of stars. These so called ``Lick indices'' have
been at the center stage of research on old stellar populations,
particularly in elliptical galaxies, for over 20 years (see Rose 2007
for a recent review). The sensitivity of these indices to fundamental
stellar parameters ($T_{eff}$, $\log g$ and $Z$) was calibrated and
models were constructed which map their behavior in terms of the $t$
and $Z$ of SSPs. Another important set of indices was defined by Bica
\& Alloin (1986) on the basis of star-cluster data of resolution
similar to the Lick library. Star clusters of all ages were included,
which allowed Bica (1988) to apply his synthesis method to late type
galaxies as well as ellipticals. (Again, see Ahumada's contribution
elsewhere in this volume for updates on this line of work.)

In the past few years, the publication of medium and high resolution,
wide $\lambda$-coverage libraries of stars across the HR diagram and
the implementation of such libraries into evolutionary synthesis codes
allowed the first attempts to model galaxy spectra on a pixel-by-pixel
basis. This represented a ``phase-transition'' in the field, a
transition which is quickly being consolidated with the ongoing
release of new and more complete libraries. While indices will still
be around for years to come, there is a very strong push towards full
spectral fits.

Before fitting a spectrum, you have to decide whether the flux
calibrations of your data and the basis spectra are reliable. If they
are not, or if for whatever reason (say, you would like to dribble
extinction issues) you do not care about the continuum, you can still
fit a {\it rectified} spectrum.

\subsection{Method}

Once you've defined a parameter space (\S\ref{sec:basis}) and your
favorite observables (\S\ref{sec:observables}), it's time to figure a
way from going from one to the other, ie, to estimate your chosen
parameters given your data. So many methods to do this journey have
been proposed that I will not dare to even briefly review it (see Cid
Fernandes 2007 for a review). Suffice it to say that this is not an
easy task. If your parameter space is of low dimensionality, life may
be easier, since you have already simplified your problem beforehand
with your own preconceptions (``priors'', in the Bayesian lingo). For
instance, you may want to model a whole galaxy as a single population,
in which case your task is to find the best $t$ and $Z$ (and perhaps
$\alpha$/Fe, if your basis allow for it), and relatively
straight-forward fitting techniques can be used. This would be
essentially equation \ref{eq:synt} with $N_\star = 1$. Ellipticals are
often modeled in this fashion.  More complex/realistic models are
obviously desirable, particularly for spirals, where the SFH is surely
not even remotely like a single burst. The larger the parameter space
(i.e., larger $N_\star$), the harder it is to explore it, and the more
degenerate the problem becomes.

{\it Degeneracies} are a recurrent theme in this business, and a
source of considerable confusion. Methods based on indices often try
to recover more populations than observables, ie., $N_\star > N_{\rm
indices}$, and people doing this have sometimes been wrongly accused
of being able to fit anything they want. Pelat (1998), in a very
elegant paper, shows that this is not true at all. If your basis is
not right and/or your data have errors, what could in principle be a
problem with infinite exact solutions may turn out to have none! And
even if there are infinite solutions, it is possible to determine the
{\it subspace} of the parameter space which maps onto the data
space. This does not eliminate degeneracies, but it tells you what the
acceptable solutions are, and this is valuable information.  Besides
algebraic issues, astrophysical effects also conspire against unique
solutions. The most (in)famous such degeneracy is the one between $t$
and $Z$. Red and strong lined galaxies may be modeled as having old
ages or large metallicities, such that old, metal poor $\vec{x}$
components in equation \ref{eq:synt} can be swapped by younger and
higher $Z$ components without much change in $L_\lambda^{gal}$, and
vice-versa. In general, components with similar $l^{SSP}_{\lambda,j}$
spectra are hard to be distinguished. For instance, it is hopeless to
try to distinguish 4 from 5 Gyr old populations with spectra of
realistic S/N. Current estimates of the age-resolution achievable with
inversion methods place it in the $\Delta \log t \sim 0.5$--1 dex
range.

Degeneracies are thus an inescapable fact-of-life in population
synthesis, and one must learn to live with them! The simpler strategy
to deal with this is to downsize your expectations and work with a
relatively coarse description of SFHs.

This can be done in a number of ways. For instance, you may start with
the number of components which you believe you are able to recover
when fitting realistic galaxy spectra, as done in the MOPED code
(e.g., Panter \etal 2003). STECMAP (Ocvirk \etal 2006) achieves this
same general goal applying regularization techniques which impose that
the derived SFH be a smooth function of time. My own code, STARLIGHT
(Cid Fernandes \etal 2005), uses a ridiculously large $N_\star$ (of
order 100), but when it comes to describe SFHs a reduced version of
$\vec{x}$ is used, grouping components of similar properties. {\it
Compression} of the population vector may thus be carried out {\it a
priori}, on the fly, or {\it a posteriori}.

There are numerous technical and astrophysical differences between
currently existing methods to go from the data to the parameters. Some
account for extinction, some don't. Some include kinematics, others
don't. Some impose simple $Z$-$t$ relations, some use a fixed $Z$ and
others treat $Z$ and $t$ independently. Some model indices, others the
full spectrum. Some compute combinations of basis elements on the fly,
others compare the data to a large library of precomputed models. Last
but not least, there are also differences in the spectral basis. 

Given this bewildering diversity, I would not be surprised to see a
proportionately large diversity in the results. To my surprise, a test
carried out at IAU Symposium 241 showed an amazing degree of
convergence among all methods. Either we all got it right, or we are
all doing the same mistakes!  Let us suppose we are on the right track
and show some results.

\section{Results: A brief tour of spectral fits with STARLIGHT}

After this necessarily simplified introduction, this section
illustrates what kind of results can be achieved with state-of-the-art
population synthesis methods. All results below were obtained with the
code STARLIGHT described in Cid Fernandes \etal (2004, 2005) and
Mateus \etal (2006).  STARLIGHT has been successfully applied to
spectra of normal galaxies, Seyfert nuclei, LINERs, Transition
Objects, quasars, luminous infrared galaxies, and others, with data
qualities spanning the full horrible--marvelous range. Let's see what
we obtain.

\begin{figure}
  \centering
  \includegraphics[width=.5\textwidth]{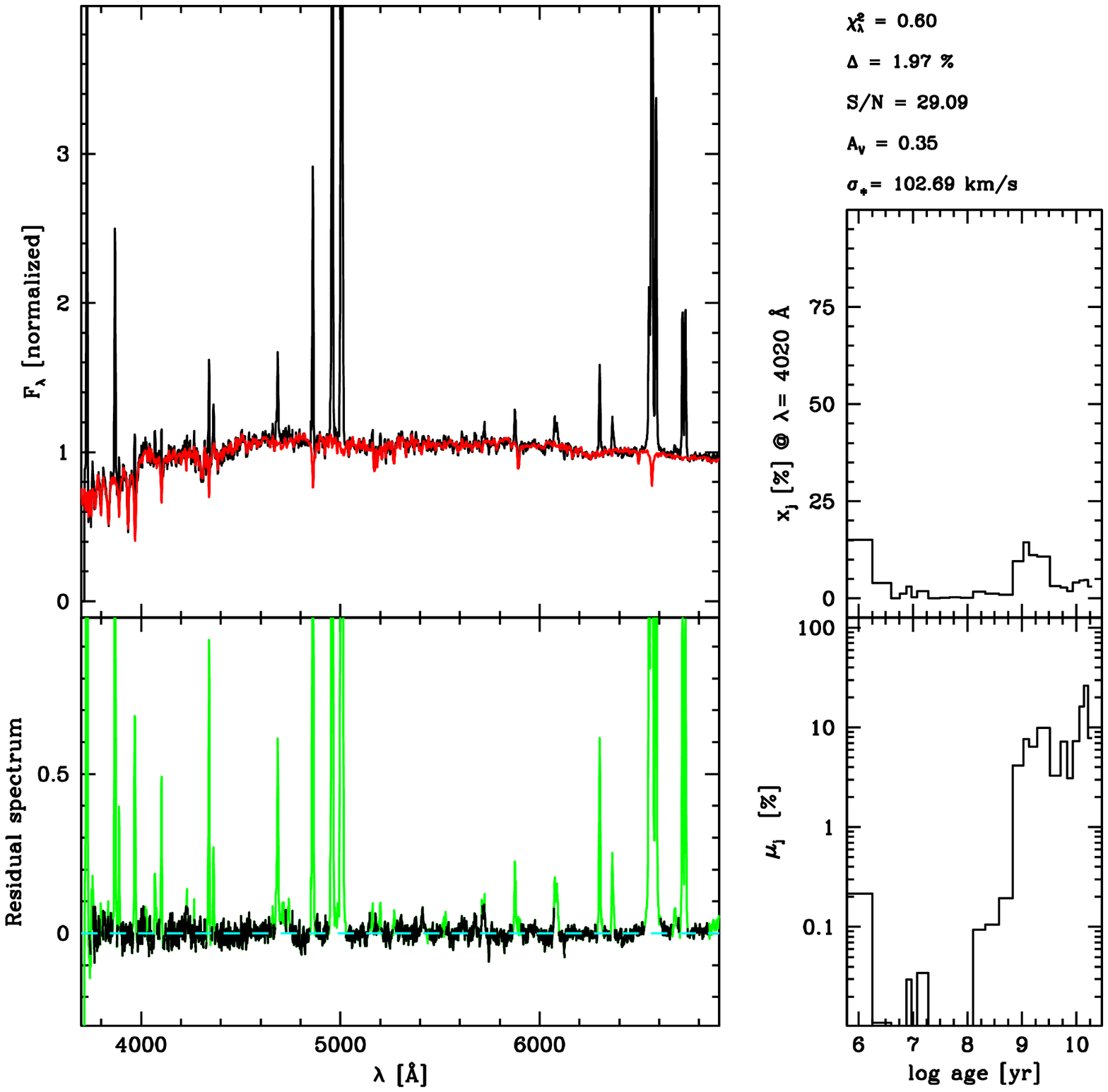}~\hfill
  \includegraphics[width=.5\textwidth]{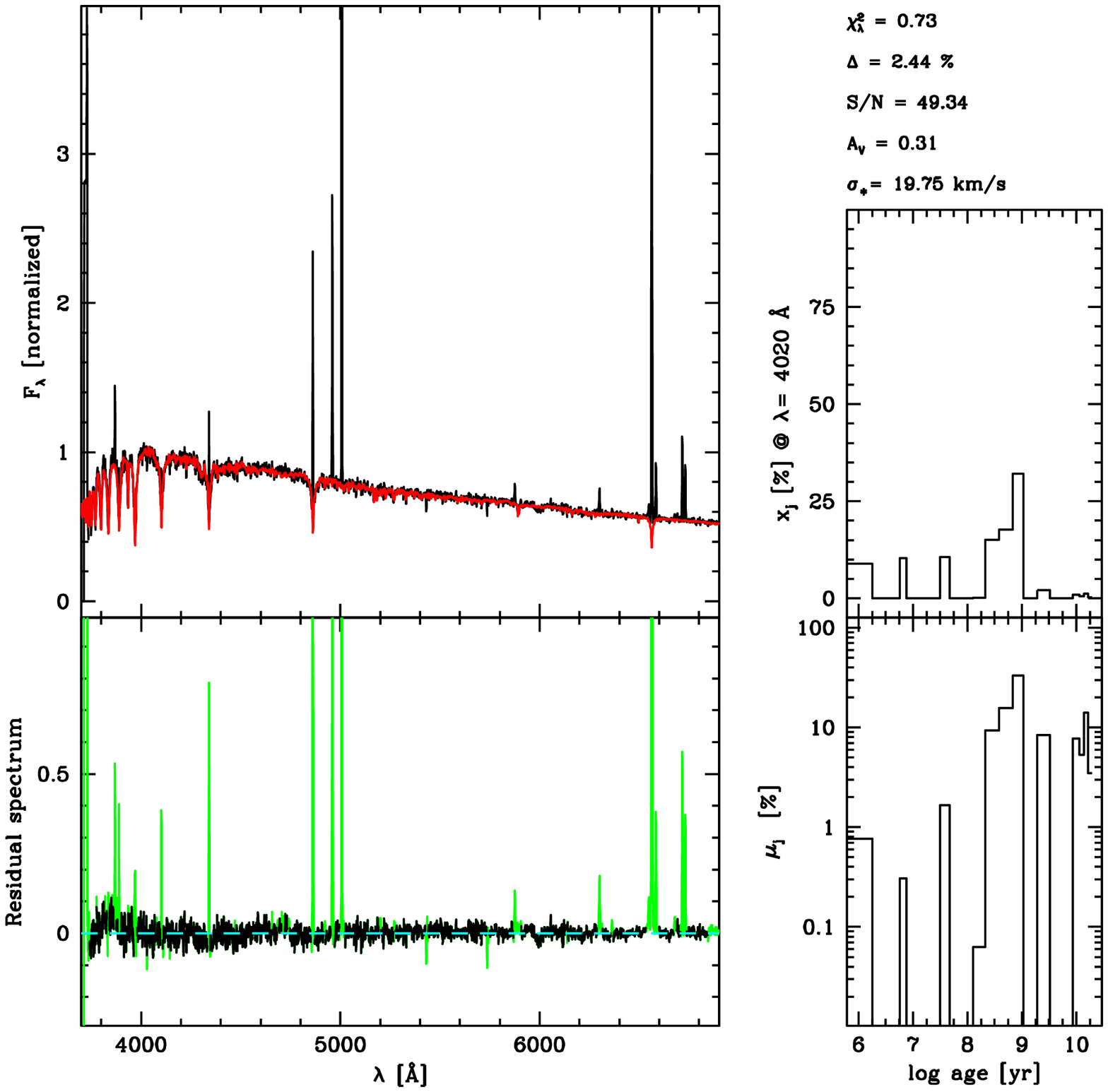}
  \caption{Examples of STARLIGHT fits to an AGN (left) and a
  star-forming galaxy (right) from the SDSS. The top panels panels
  show the observed (black) and model (red) spectra, while the bottom
  ones show the residuals, with emission lines and bad pixels are
  marked in green Green. The smaller panels show the derived age
  distribution in light (top) and mass (bottom) fractions.}
  \label{fig:Fig_SDSS_fits}
\end{figure}

Fig.~\ref{fig:Fig_SDSS_fits} shows fits to 2 of the nearly {\it
600000} SDSS galaxies which have already been modeled with
STARLIGHT. The fits were carried out with a basis of SSPs spanning 25
ages and 6 metallicities ($N_\star = 150$!) from BC03.  The left panels
plot the observed, model and difference spectra, while the right
panels show the inferred light and mass population vectors without any
compression apart from the marginalization over $Z$. Estimates of the
SFH and chemical enrichment histories, starlight extinction, velocity
dispersion and stellar masses have been obtained for all
galaxies. Emission lines measurements, for which accurate subtraction
of the stellar spectrum is critical, have also been measured, yielding
estimates of nebular metallicity, H$\alpha$/H$\beta$-extinction, line
widths and etc. 

\begin{figure}
  \centering
  \includegraphics[width=0.4\textwidth]{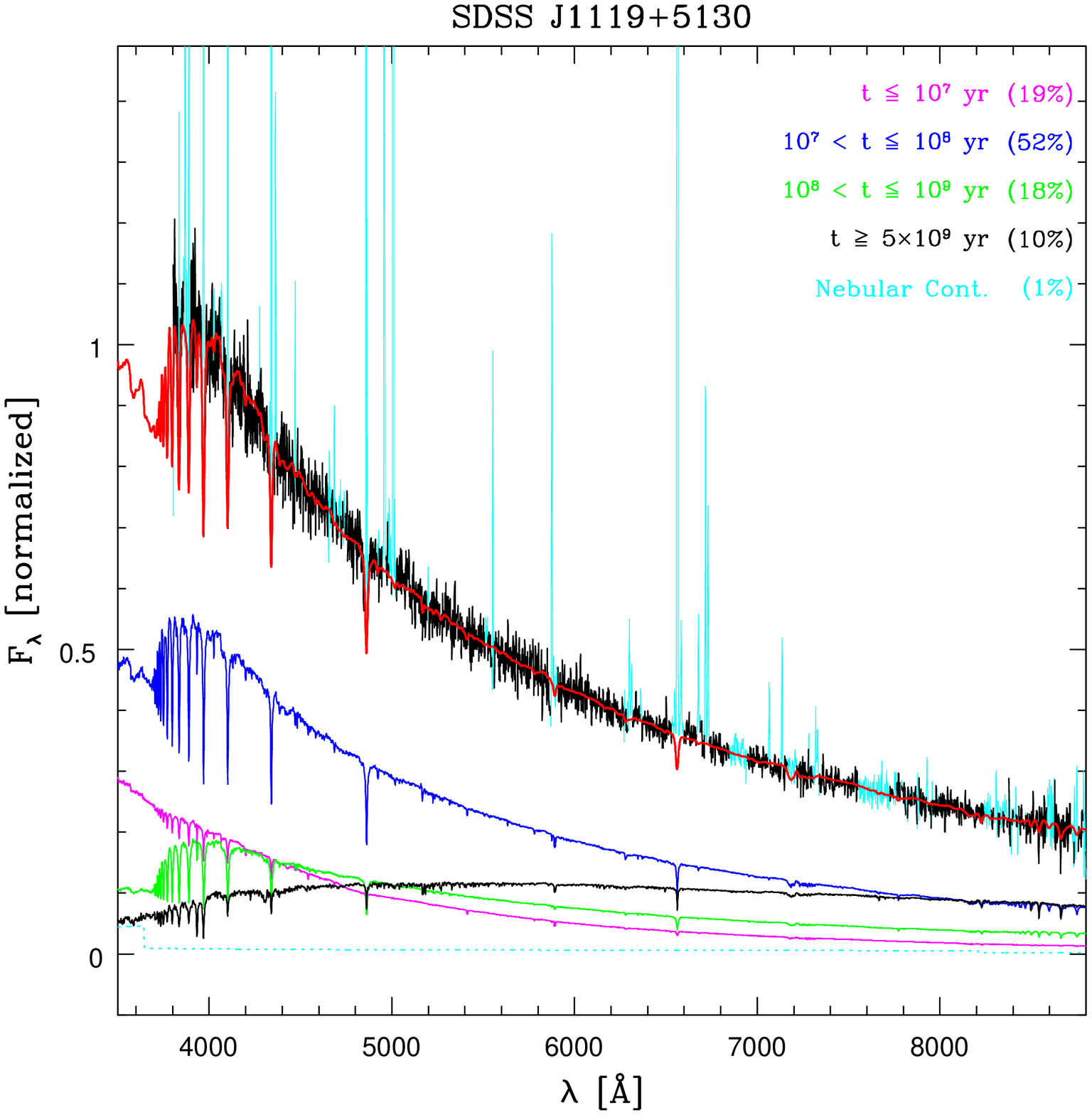}
  \includegraphics[width=.55\textwidth]{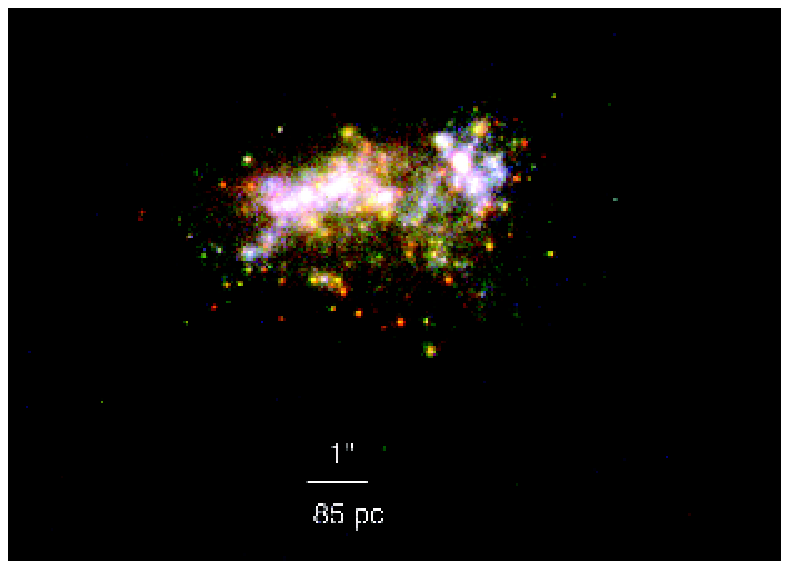}
  \caption{HST color image and STARLIGHT fit to the spectrum of SDSS
  J1119+5130. The observed and model spectra are shown in black and
  red, respectively, Emission lines and bad pixels (turquoise) were
  masked in the fits. The bottom spectra are sums of the constituent
  SSP spectra in four age bins, whose contributions to the observed
  flux at 4020 \AA\ are shown in parentheses. See Corbin \etal
  (2006) for details.}
  \label{fig:Fig_UCBDGs}
\end{figure}

Fig.~\ref{fig:Fig_UCBDGs} shows a STARLIGHT fit to the SDSS spectrum
of an Ultra Compact Blue Dwarf Galaxy (Corbin \etal 2006). 25 ages but
only $Z_\odot/50$ SSPs were included in the fits of these SMC-like
galaxies, which, despite looking extremely young, have most of their
stellar mass locked up in old, $\sim 10$ Gyr stars. As in IZw18, the
ongoing star formation performs a pretty good ``plastic surgery'' in
these systems, but not effective enough to hide its true age from the
synthesis. The plot groups the 25 components of $\vec{x}$ onto just 4
relevant age bins, illustrating our {\it a posteriori} compression
approach.

\begin{figure}
  \centering
  \psfig{file=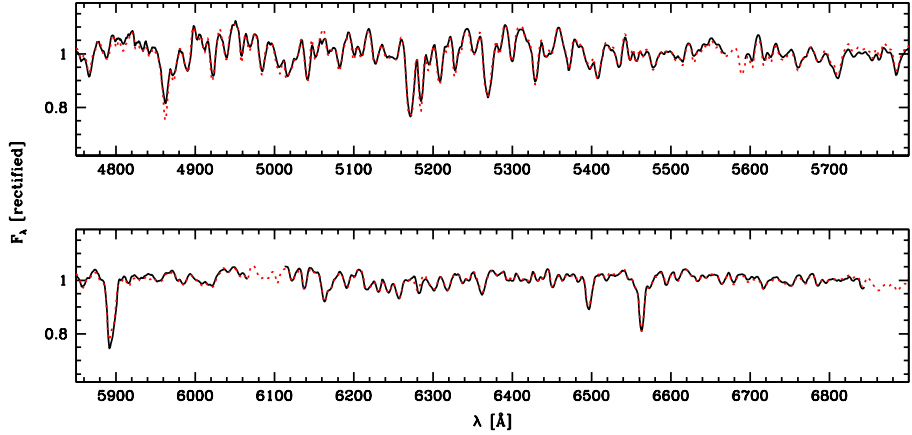,width=\textwidth,bbllx=20,bblly=430,bburx=600,bbury=710}
  \caption{Section of a high $S/N$ Gemini spectrum modeled with
  STARLIGHT. Observed and model spectra are shown in black (solid) and
  red (dotted) respectively (Coelho \etal 2007, in prep.).}
  \label{fig:Fig_M32}
\end{figure}

The astute reader may be wondering whether the success stories told by
Figs.~\ref{fig:Fig_SDSS_fits} and \ref{fig:Fig_UCBDGs} are just due to
the fact that we are fitting relatively noisy data.  While it is true
that fitting good data is always harder, Fig.~\ref{fig:Fig_M32} shows
that STARLIGHT is also capable of dealing with $S/N \sim 100$
spectra. (This exquisite spectrum was obtained by Paula Coelho and
Claudia Mendes de Oliveira.)  Data and model can hardly be told apart!

The message from this quick tour is that one can nowadays model galaxy
spectra on a pixel-by-pixel basis with an unprecedented degree of
success. Though all examples are based on STARLIGHT, I am sure that
MOPED, STECMAP, and other ``competing'' codes would perform equally
well, which lends credibility to spectral fits.

\section{Checks, caveats and final words}

By themselves, spectral fits may look good and be very helpful to
derive pure-emission spectra. Their main utility, however, is to derive
the SFH of galaxies. Many interesting results have emerged in
recent years, both using just the first moments of the $t$ and $Z$
distributions and the full SFHs.  Instead of reviewing these
astrophysical results (see Cid Fernandes 2007 and references therein),
I prefer to end this contribution illustrating a couple of the many
sanity checks which we have carried out, and an illustration of how
things which go wrong can also teach us important lessons.

\begin{figure}
  \centering
  \includegraphics[width=0.45\textwidth,height=0.35\textwidth]{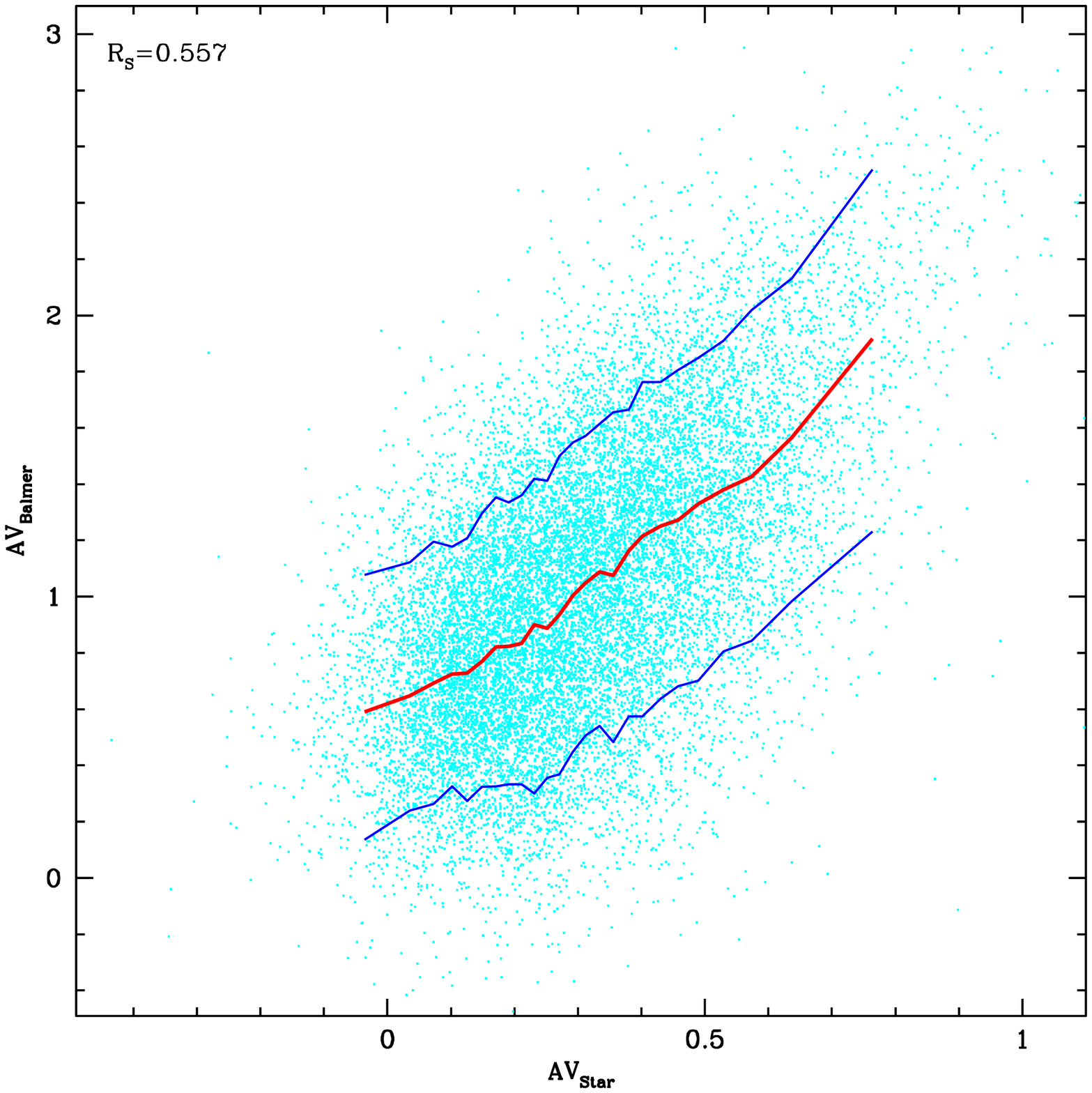}
  \includegraphics[width=0.45\textwidth,height=0.35\textwidth]{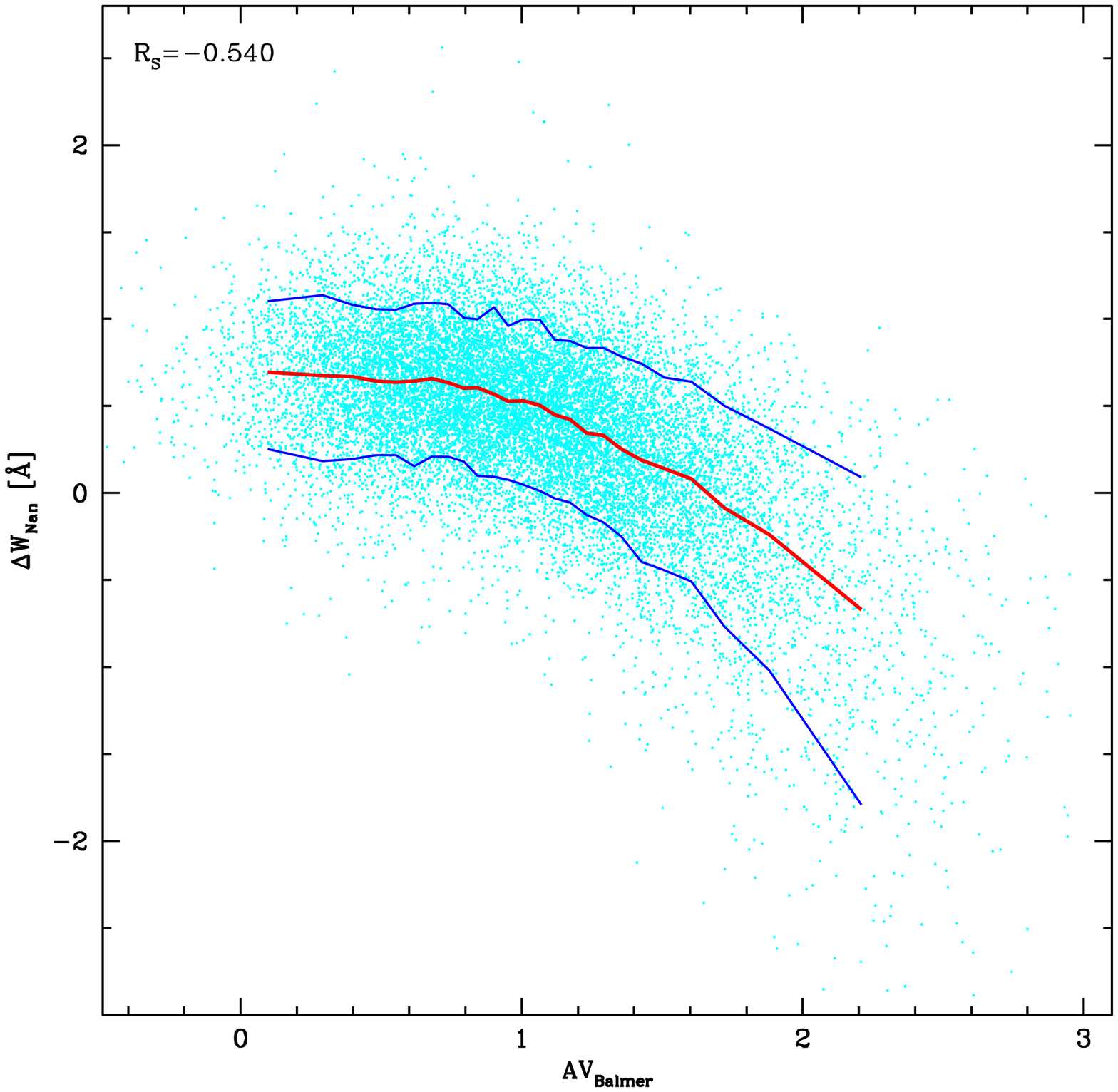}
  \caption{{\it Left:} Nebular versus stellar extinction for high
  $S/N$ star-forming galaxies in the SDSS. {\it Right:} Equivalent
  width of the NaD $\lambda\lambda$5890,5896 doublet, measured from
  the residual spectrum, against the nebular extinction derived from
  H$\alpha$/H$\beta$ for star-forming galaxies. Lines mark the 5, 50
  and 95\% percentiles.}
  \label{fig:Fig_NaD}
\end{figure}

As already mentioned, naive modeling of dust effects is perhaps the
major caveat in spectral fits. To check whether the fits are producing
at least sensible results, we have correlated the derived values of
$A_V$ with those derived from the H$\alpha$/H$\beta$ ratio.
Fig.~\ref{fig:Fig_NaD} shows that these two independently quantities
correlate very strongly. Not only that, we find that stars suffer
about half the extinction that affects the gas, in amazing agreement
with detailed studies of nearby star-forming galaxies (Calzetti \etal
1994). Fig.~\ref{fig:Fig_NaD} shows a further test of the consistency
of the synthesis results. We correlate the nebular extinction,
$A_V(H\alpha/H\beta)$, with the strength of the residual NaD
absorption doublet, which traces the cold ISM component. The
gratifyingly evident relation between these two independently derived
quantities gives us confidence that our results are meaningful,
despite the over-simplified dust geometry adopted. Cid Fernandes \etal
(2005) and Asari \etal (2007, in prep.) present further sanity checks.

\begin{figure}
  \centering
  \psfig{file=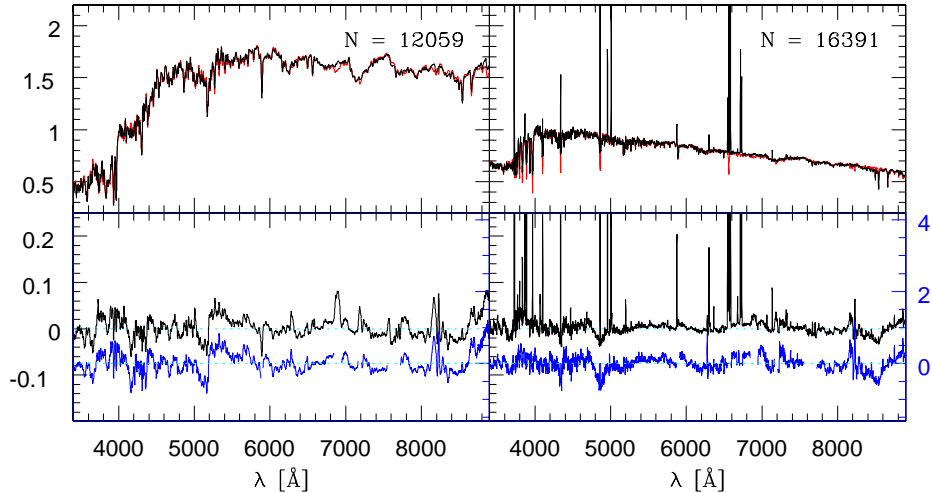,width=\textwidth,bbllx=20,bblly=400,bburx=600,bbury=710}
  \caption{ {\it Top:} Mean observed spectra (black) and their
respective models (red) for thousands of high $S/N$ elliptical (left)
and star-forming (right) SDSS galaxies, all normalized at 4020
\AA. {\it Bottom:} Zoom of the Observed $-$ Model (black) and
(Observed $-$ Model) / Error (blue) residual spectra.}
  \label{fig:Fig_ResidSpectra}
\end{figure}

Having spent so many lines emphasizing how great spectral fits are, it
is fitting to warn you that they are not
perfect. Fig.~\ref{fig:Fig_ResidSpectra} illustrates this. What is
shown there are the stacked results of fits for many thousands of high
$S/N$ elliptical (left) and late-type, star-forming galaxies (right)
from the SDSS. The top panels show the mean observed and model
spectra. Seen in their natural scales, the fits look perfect. However,
zooming on the the mean {\it residual spectra} (bottom panels) reveals
several systematically miss-fit spectral features.

In elliptical galaxies, features associated to $\alpha$ elements, like
the Mg, CN and Na bands, are all under-fitted. This is not surprising,
given that the library used by BC03 is fundamentally based on solar
neighborhood stars, whose chemical abundance pattern are known to
differ from stars in early type systems.  In star-forming galaxies,
selected on the basis of emission line diagnostic-diagrams, the most
obvious problem is a shallow but broad ``absorption'' band around
H$\beta$, which appears whenever stars of $\sim 100$ Myr are present
in large proportions.  Unlike the problems for $\alpha$-related bands
in ellipticals, there is no straightforward physical explanation nor
element identification for this trough. We suspect that this is not an
absorption band at all, but a side-effect of flux calibrations issues
in STELIB, maybe associated to the fact that stars with wide H$\beta$
absorption were used as flux standards. In any case, these miss-fits
illustrate how the analysis of galaxy spectra can provide useful
feedback to those constructing stellar and SSP spectral libraries. It
will be important to repeat this practical-tests with the new
libraries to evaluate whether they do bring the expected improvements.

To close, I hope these pages have given you an idea of the several
choices one has to make when modelling galaxy spectra, and how this
technique to perform ``paleontological'' studies of galaxies has
matured substantially in the past few years due to a combination of
advances in theoretical ingredients, mathematical methods and
abundance of data. Further progresses in these 3 fronts are expected
in the very near future. Clearly, we will have busy and exciting times
ahead.

\end{document}